\documentstyle[aps,preprint]{revtex}
\tighten
\begin{document}
\draft
\title{\bf Fractional Quantum Hall Effect Measurements at Zero g-Factor}

\author{$^{1}$D.R.~Leadley, $^{2}$R.J.~Nicholas, $^{3}$D.K.~Maude, $^{4}$A.N.~Utjuzh, 
$^{3}$J.C.~Portal, $^{5}$J.J.~Harris and $^{6}$C.T.~Foxon}

\address{$^{1}$ Department of Physics, University of Warwick, Coventry, CV4~7AL,~UK\\
$^{2}$Department of Physics, Clarendon Laboratory, Parks Road, Oxford, OX1~3PU,~UK\\
$^{3}$Grenoble High Magnetic Field Laboratory, MPI-CNRS, 25~Avenue des Martyrs 
BP~166, F-38042~Grenoble Cedex~9, France\\
$^{4}$Russian Academy of Sciences, Institute of High Pressure Physics, 142092~Troitsk, 
Moscow Region, Russia\\
$^{5}$Department of Electronic and Electrical Enginering, University College, London, 
WC1E~7JE,~UK\\
$^{6}$Department of Physics, Nottingham University, University Park, Nottingham, 
NG7~2RD,~UK}

\date{Submitted to Physical Review Letters, \today}

\maketitle

\begin{abstract}

Fractional quantum Hall effect energy gaps have been measured in 
GaAs/Ga$_{0.7}$Al$_{0.3}$As heterojunctions as a function of Zeeman energy, which is varied 
by applying hydrostatic pressure up to 20~kbar. The gap at $\nu=1/3$ decreases with pressure 
until the g-factor changes sign when it again increases. The behavior is similar to that seen at 
$\nu=1$ and shows that excitations from the 1/3 ground state can be spin-like in character. At 
small Zeeman energy, the excitation appears to consist of 3 spins and may be interpreted as a 
small composite skyrmion.
\end{abstract}

%\narrowtext

The two dimensional electron gas in a high magnetic field is an excellent test bed for studying 
electron-electron interactions.  In recent years our understanding has been greatly simplified by 
the composite Fermion (CF) model, which maps the fractional quantum Hall effect (FQHE) of 
electrons onto an integer quantum Hall effect (IQHE) of CFs \cite{fqhe,cft,cfe}. Thus the physics 
of the state at filling factor $\nu=1/3$, where there is one completely occupied CF Landau level 
(LL), is explained by analogy with $\nu=1$. The other principal FQHE states at $\nu=p/(2p+1)$ 
can similarly be explained by the integer states at $\nu=p$. While the ground states are quite well 
understood the same is not true of the excited states which are responsible for conduction when 
the Fermi energy lies in a mobility gap.

The state at $\nu=1$ is an itinerant ferromagnet with a spontaneous magnetization, and 
consequently the activation energy gap deduced from transport measurements was found to be 
much larger than the single particle Zeeman energy (ZE).  Recently it has been shown optically 
\cite{nmr} and electrically \cite{duncan,nu1,eise} that the excitations at this point are probably 
spin texture excitations which in the limit of vanishing ZE are skyrmions \cite{ferr,sond}.  In this 
paper we examine the CF analogue of this state at  $\nu=1/3$ as the ZE vanishes.  Our 
measurements suggest that in this limit a new skyrmionic CF excitation occurs.

Although the initial CF model ignored spin it is very important when the Land\'e $g$-factor is 
small. In GaAs the electronic ZE $g\mu_BB$ has similar magnitude to the gaps between CF LLs, 
which arise from electron-electron correlations and scale like $E_c=e^2/4\pi\epsilon l_B$ 
($l_B=\sqrt{\hbar/eB}$ is the magnetic length). CF LLs originate from $\nu=1/2$ where there is 
an offset of $g\mu_BB_{1/2}$ between fans of each spin, which provides an essential difference 
from the IQHE. This gives the possibility of level crossing as the ZE and magnetic field are varied 
and leads to the observed disappearance and re-emergence of fractions \cite{dl32,du32,tilt}. 
Although the ground state at $\nu=1/3$ will be fully spin polarized, the states at $\nu=2/3$ or 2/5 
may be either fully polarized or unpolarized depending on the relative sizes of the ZE and the CF 
LL gaps. Similarly the excitations may involve either spin flips or inter CF LL transitions. At 
$\nu=1/3$ and small ZE, i.e.\ very low magnetic fields or small g-factor, we expect a spin flip 
transition to the lowest CF LL state with the opposite spin. The interesting question which we 
address is whether this is a single spin flip of one CF or a collective phenomenon i.e.\ a 
skyrmionic excitation of the CFs which we will refer to as a composite skyrmion.

We have performed experiments where $g$ can be tuned through zero thus favoring skyrmion 
formation. The tuning is achieved with hydrostatic pressure of up to 22~kbar \cite{pres}. In 
GaAs $g=-0.44$, as a result of subtracting band structure effects from the free electron value of 
2. At higher pressure the band structure contribution reduces, and so does the magnitude of $g$ 
which passes through zero at $\sim$ 18~kbar. Previously we used this method to investigate the 
changing energy gaps of the mixed spin states around $\nu=3/2$ \cite{dl32}. Here we 
demonstrate that the gap at $\nu=1/3$ is indeed a spin gap with excitations consistent with 
flipping $\sim$3 spins at small ZE. This shows that composite skyrmions can be formed at 
$\nu=1/3$ when the electron g-factor is sufficiently small. By contrast the gap at $\nu=2/5$ is 
consistent with a single particle excitation.

The samples studied were high quality GaAs/ Ga$_{0.7}$Al$_{0.3}$As heterojunctions grown 
by Molecular Beam Epitaxy at Philips Research Laboratories, Redhill. Samples G586, G627 and 
G902 have undoped spacer layers of 40, 40 and 20~nm. At ambient pressure and 4~K their 
respective electron densities after photoexcitation are 3.3, 3.5 and $5.7\times 10^{15}{\rm 
m}^{-2}$ with corresponding mobilities of 300, 370 and 200~m$^2$/Vs. Data from similar 
samples measured without applied pressure is included from Ref.~\cite{prb}. The samples were 
mounted inside a non-magnetic beryllium copper clamp cell \cite{cell} and the pressure was 
measured from the resistance change of manganin wire. The absolute values quoted at low 
temperature are accurate to $\pm1$~kbar, but between data points the variation is less than 
$\pm0.2$~kbar. The pressure cell was attached to a top loading dilution refrigerator probe 
allowing temperatures as low as 30~mK to be obtained and measured with a ruthenium oxide 
resistor attached outside the pressure cell, which followed the sample temperature with a 
negligible time lag.

Increasing the pressure causes the GaAlAs conduction band to move relative to the GaAs 
conduction band in the well reducing the number of electrons. Above $\sim 13$~kbar no 
electrons were present in the dark at low temperature, but a certain number could be recovered 
after illumination from a red LED. The illumination time required to obtain a constant number of 
electrons roughly doubled for every 2~kbar increase in pressure, reaching several hours at 
20~kbar. The highest pressure studied was 22~kbar, but no conductivity could be measured 
despite prolonged illumination. The sample required several hours for the density to stabilize 
before quantitative measurements could be made during which it varied by less than 1\% over the 
full temperature range. The data from G586 was recorded with a density of $0.44\pm0.06\times 
10^{15}{\rm m}^{-2}$ above 13~kbar and slightly higher at lower pressures. This puts 
$\nu=1/3$ at 5.4~T. For G627 and G902 the data was recorded in the range 0.77--1.23$\times 
10^{15}{\rm m}^{-2}$ i.e.\ $\nu=1/3$ at 9--15T.

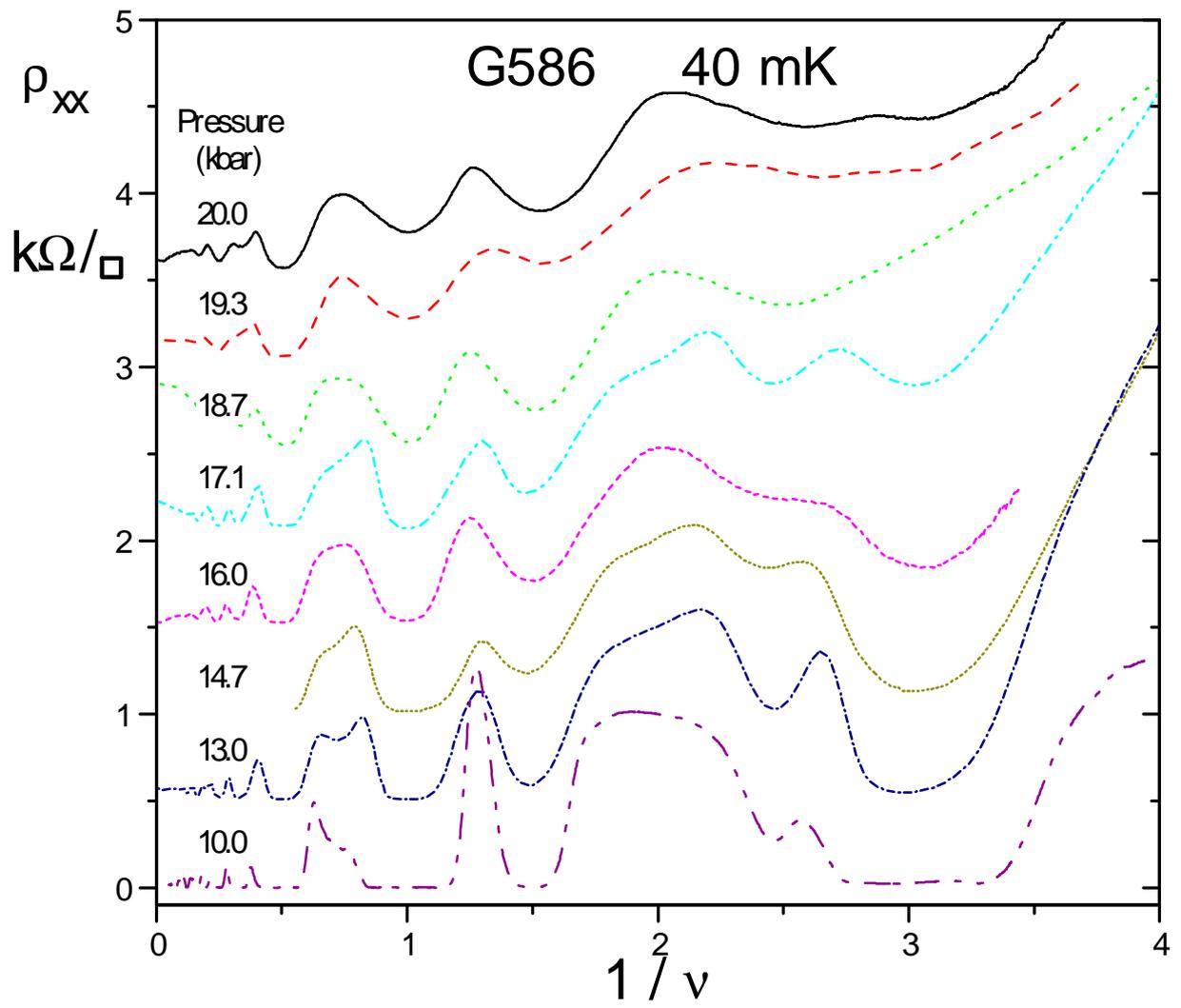
\begin{figure}
\caption{Magnetoresistance recorded at 40~mK for sample G586 at pressures between 10 and 
20~kbar. Note how the feature at $\nu=1/3$ becomes weaker relative to 2/5 with increasing 
pressure, but is recovered at the highest pressure.}
\label{fig:rxx}
\end{figure}

The magnetoresistance $\rho_{xx}$ of sample G586 at 40~mK is shown for pressures between 
10 and 20~kbar in Fig.~\ref{fig:rxx}, plotted against $1/\nu$ to remove the remaining small 
density variation. The feature at $\nu=1/3$ weakens as the pressure is increased, it completely 
disappears at 18.7~kbar and is recovered at the highest pressure. Meanwhile, the feature at 
$\nu=2/3$ remains approximately constant, which is an important indication that pressure does 
not denigrate the sample quality and destroy the FQHE.

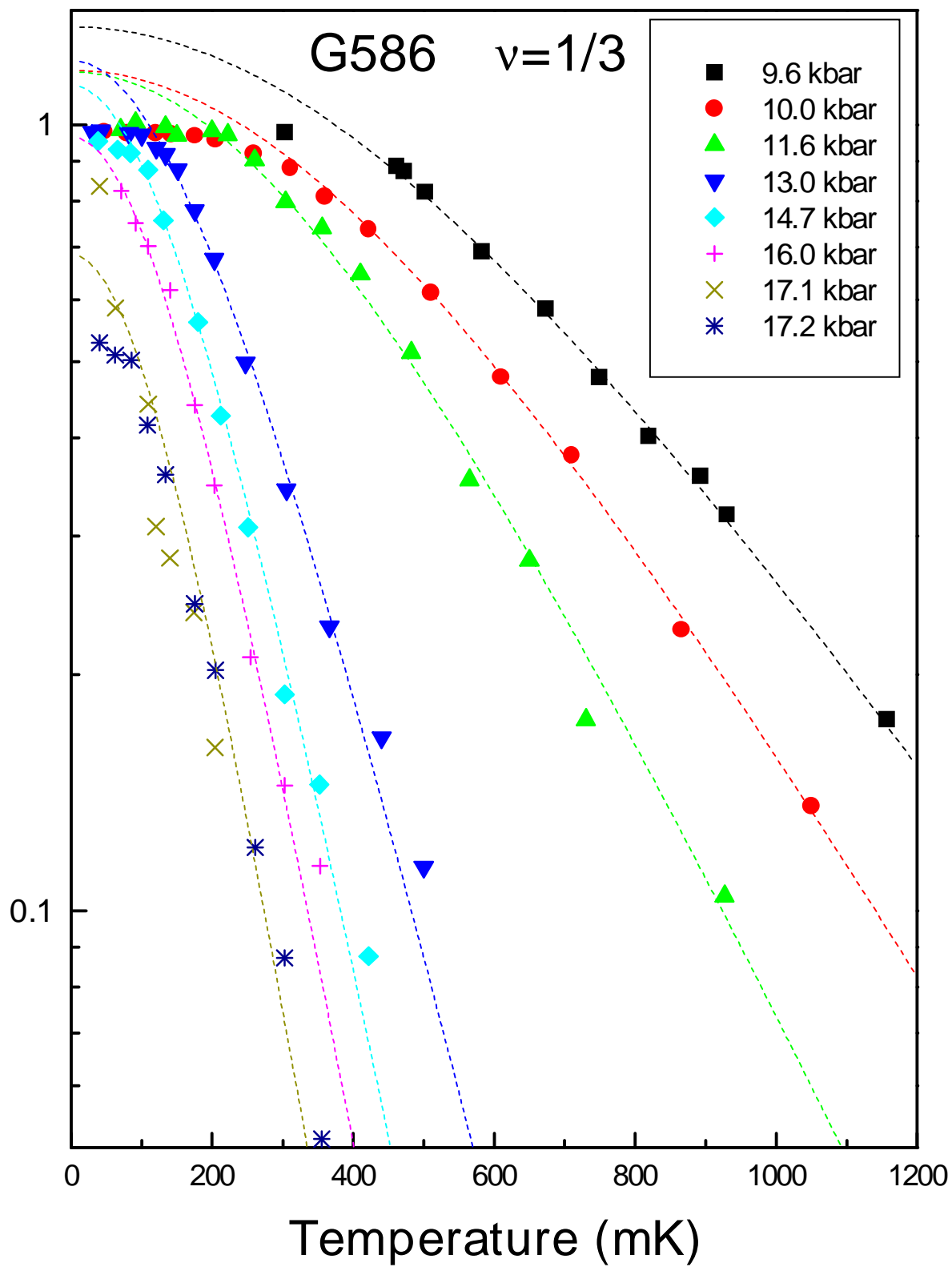
\begin{figure}
\caption{Temperature dependence of the $\nu=1/3$ minimum, $\Delta\rho_{xx}/\rho_{xx}$ for 
sample G586. The energy gap is extracted from fits to the LK formula shown by dashed lines.}
\label{fig:ando}
\end{figure}

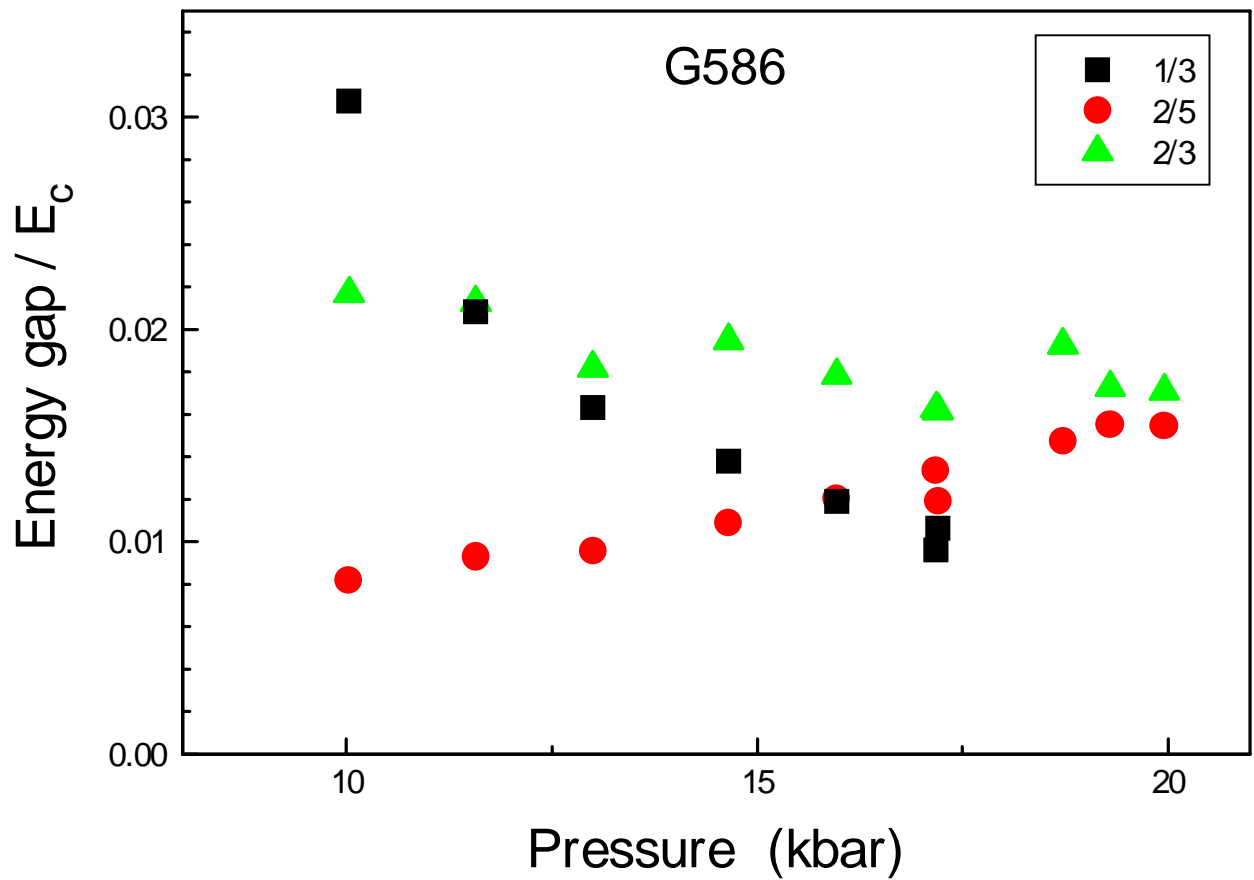
\begin{figure}
\caption{Energy gaps measured for the strongest fractions in sample G586, showing how the gap 
at $\nu=1/3$ decreases, 2/5 increases and 2/3 remains approximately constant as the applied 
hydrostatic pressure is increased. Above 19~kbar the gap at 1/3 increases again but reliable 
quantitative values can not be obtained. }
\label{fig:evp}
\end{figure}

Figure~\ref{fig:ando} shows the temperature and pressure variation of the 1/3 minimum, defined 
as $(\rho_{xx}(\infty)-\rho_{xx}(T))/\rho_{xx}(\infty)$, where $\rho_{xx}(\infty)$ is the 
resistivity at the same field taken from a high temperature trace where there is no longer a 
minimum. From the figure it is clear that at higher pressures progressively lower temperatures are 
required to see a 1/3 minimum. Thus, however the data is analyzed, the energy gap $E_g$ will 
decrease strongly with pressure. We have extracted values of $E_g$ by fitting the temperature 
dependence to the Liftshitz-Kosevich (LK) formula, from which $\Delta\rho_{xx}\propto X/\sinh 
X$ where $X=2\pi^2kT/E_g$. This procedure, described in more detail in Ref.~\cite{prb}, has 
the advantages over finding activation energies from an Arrhenius plot that firstly it measures the 
gap between LL centers not the mobility gap, and so is less sensitive to changes in disorder, and 
secondly an accurate zero of resistance is not required, which avoids any problems of parallel 
conduction and means especially low temperatures are not required. Values of $E_g$ at 
$\nu=1/3$, 2/3, and 2/5 are shown in Fig.~\ref{fig:evp} in units of $E_c$ for G586. In these 
units a gap of 0.01 is equivalent to 1.17~K for $\nu=1/3$ at 5.4~T and 0.83~K for $\nu=2/3$ at 
2.7~T. This scaled data shows the same trends as the raw data, but the scatter due to the small 
density variation between different pressures is removed. Scaling the data also makes comparison 
with theory easier.

The trends observed in the raw data can now be quantified and it is seen that 1/3 decreases and 
2/5 increases with pressure over this range. Experimentally the feature at 1/3 vanishes between 
17 and 19~kbar, which is exactly the pressure region where $g$ is predicted to pass through 
zero. By vanishing we mean that 1/3 is weaker than 2/5 and a separate minimum cannot be 
observed, although the 2/5 minimum has a pronounced tail on the high field side from the residual 
1/3 feature. Thus an upper limit can be set on the 1/3 gap, although we can not tell if it has 
completely collapsed. Even though a 1/3 feature could be seen at 20~kbar in the lowest 
temperature data, it was not possible to obtain an accurate value for the energy gap as the 
minimum could not be followed to higher temperatures. Taking the temperature dependence of 
$\rho_{xx}$ at the field where 1/3 occurs at base temperature, an energy gap of $0.017E_c $ 
results which is consistent with the fraction being established again once $g$ has changed sign.

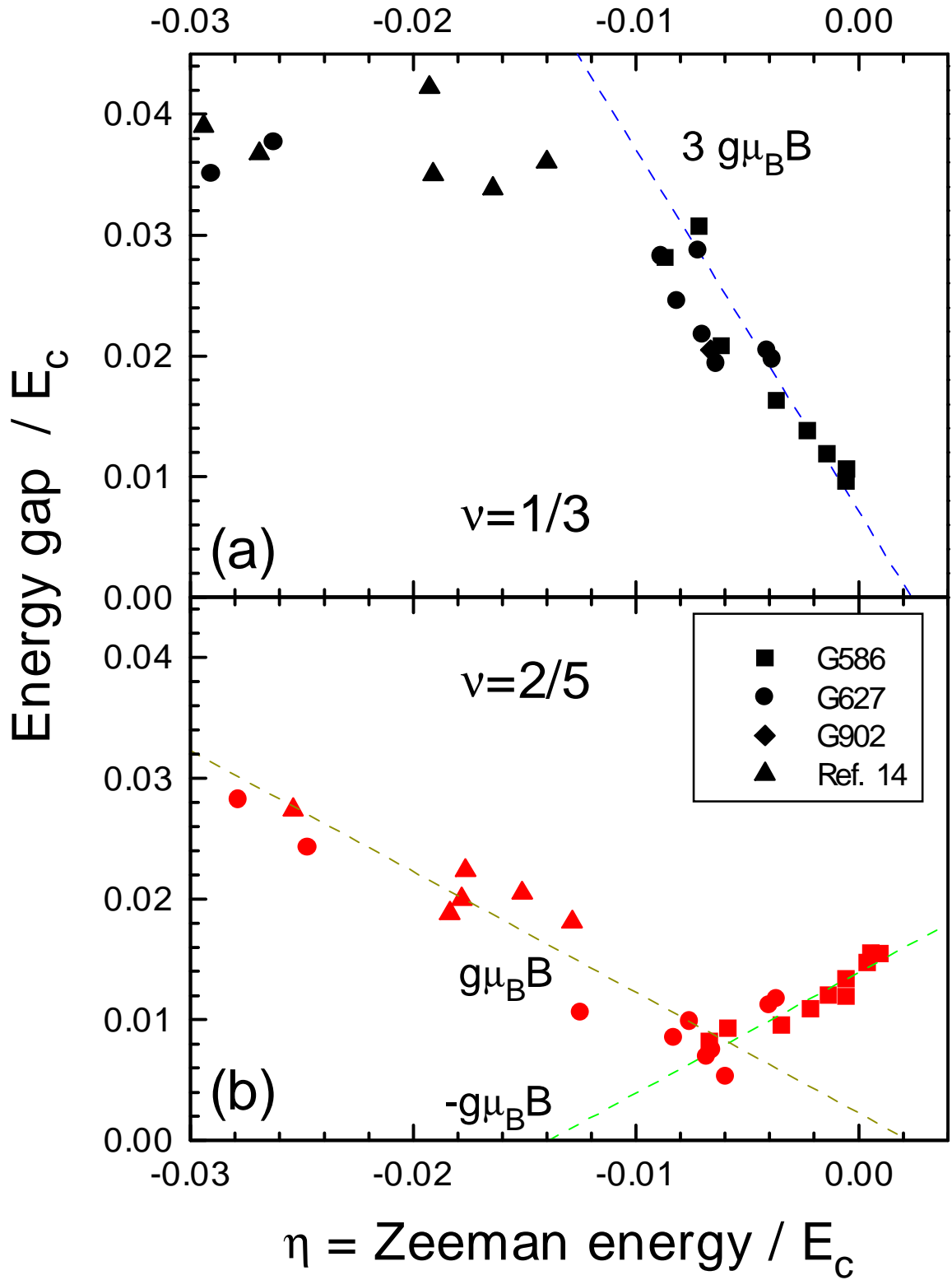
\begin{figure}
\caption{(a) Energy gap at $\nu=1/3$ for all the samples studied as a function of the Zeeman 
energy, (both in units of $E_c$). The line shows the energy required to flip 3 spins. (b) The 
energy gap at $\nu=2/5$. The slope of the lines now corresponds to a single spin flip.}
\label{fig:e13}
\end{figure}

As $g$ varies with both pressure and density, the ZE must also be scaled to compare data from 
different samples. Figure~\ref{fig:e13} shows the energy gaps at 1/3 and 2/5 for all the samples 
studied as a function of ZE. Both axes are scaled by $E_c$ so the x-axis is 
$\eta=g\mu_BB/E_c$, the ratio which determines the skyrmion size and energy \cite{eta}.

Considering 1/3 first, the data falls into two distinct groups.  For $\eta>0.01$, mostly from data 
taken at ambient pressure, the gap only scales with the Coulomb energy. This behavior is very 
similar to that observed at $\nu=1$ \cite{duncan,nu1} and shows the FQHE state at $\nu=1/3$ 
has a Coulomb gap, which may correspond to either the spin-wave or more probably the CF gap.  
For $\eta<0.01$, using data taken above 9~kbar, there is a spin gap proportional to ZE. The line 
on Fig.~\ref{fig:e13}(a), with a gradient of 3, fits the data very well at small $\eta$. A slope of 
unity  cannot account for data at small ZE. The slope of 3 corresponds to an energy gap of 
$3g\mu_BB$ which indicates an excitation involving the reversal of three spins.

This excitation could be a small composite skyrmion, as predicted by theory. In a rough estimate 
Sondhi~{\em et al.} suggested that a skyrmion formed at $\nu=1/3$ occurring at 1~T should 
contain `a couple of reversed spins' \cite{sond}. They also estimate the skyrmion--antiskyrmion 
pair gap as $0.024E_c $ at $g=0$. The minimum gap we obtain is $0.01E_c$, which compares 
well when account is taken of the typical 50\% reduction in Coulomb energies found in 
calculations where finite thickness is included \cite{das}. In a more detailed calculation the 
energy to creating an antiskyrmion at $\nu=1/3$, i.e.\ the energy to remove one spin at fixed 
magnetic field, was found to be 
$E_{1/3}/E_c=0.069+0.024 \exp\left(-0.38 R^{0.72}\right)+\eta R$ \cite{kwj}.
The number of reversed spins $R$ in the composite skyrmion can be found by minimising this 
expression and we see that $R=1$ for $\eta>0.004$; $R=3$ at $\eta=0.002$ and $R=6$ at 
$\eta=0.001$. These numbers cannot be directly compared with our experiments at fixed particle 
number where the excitation is a skyrmion--antiskyrmion pair because they do not include 
creation of the quasi-particle skyrmion or finite thickness effects. Nonetheless, they allow us to 
estimate relevant energy and size scales. It is clear that composite skyrmions will always be small 
for experimentally accessible parameters and that a size of 3 spins provides good agreement 
between the experiment and theory in the region of $\eta=0.002$. The experiment suggests 
however that the minimum gap for skyrmionic CF excitations is much less than half of the gap at 
large ZE. This is substantially different from the prediction of exactly a 50\% reduction in the gap 
at $\nu=1$ made in Ref.~\cite{sond} for infinite sized skyrmions.

Turning now to the full set of scaled data obtained at $\nu=2/5$ (Fig.~\ref{fig:e13}(b)), there are 
two distinct regions that cross over at $\eta\simeq -0.006$. For $\eta<-0.006$ the gap decreases 
as the size of the spin splitting decreases and for $\eta>-0.006$ the gap increases again. This 
suggests a level crossing and finds a straight forward explanation in the CF picture. The 
$\nu=2/5$ FQHE gap occurs when two CF LLs are full. When the ZE is small these will be the 
lowest LLs of the two opposite spin ladders, thus the excitation at $\nu=2/5$ is a spin flip from 
an unpolarized ground state. As the ZE increases the spin reversed ladder moves up relative to 
the other spin and the 2/5 gap decreases. When the second and third levels cross over there is a 
transition to a fully polarized ferromagnetic ground state and the gap might be expected to 
vanish.  Further increase of ZE opens the $\nu=2/5$ gap again, until the spin flip is no longer the 
lowest excitation and the gap saturates at $\hbar\omega_c^*$. The slope of unity observed on 
Fig.~\ref{fig:e13}(b) shows that this description in terms of single particle CF energy levels is 
valid except in the immediate neighborhood of $\eta=-0.006$, where a finite gap remains. For 
$kl_B=0$ excitations the cross over would be expected when $g\mu_BB=\hbar\omega_c^*$, 
which is clearly not the case since, from the gap saturation value and our previous work 
\cite{prb}, $\hbar\omega_c^*$ is $\sim 0.03E_c$ at $\nu=2/5$. However, the large $kl_B$ 
excitations, that transport experiments measure, will again be spin waves with a much greater 
energy than $g\mu_BB$, causing the cross over to occur at a smaller value of $\eta$. Thus there 
will be an anti-crossing of the levels and a finite gap between the ground and excited states of the 
ferromagnet formed at $\nu=2/5$. The formation and excitations out of this state would be very 
interesting to study theoretically.

While the gap at $\nu=2/3$ appears to be approximately constant over the range of pressure in 
Fig.~\ref{fig:evp}, the field at 2/3 is only half that at 1/3 which makes the range of ZE 
insufficient to draw definitive conclusions. When the scaled data at large ZE from other samples 
with $\eta<-0.01$ is included, the gap decreases by $g\mu_BB$ in a manner similar to 2/5. In the 
CF model 2/3 and 2/5 are expected to behave in a very similar way as they both have the same 
CF LLs structure. We do not see an obvious minimum in the region $-0.01<\eta<0$, although the 
scatter is larger than for 2/5.  While we can not see the levels cross over for 2/3 this has 
previously been observed when the Zeeman energy is increased by tilting the magnetic field, but 
only for the lowest density samples \cite{tilt2}. Interestingly the tilted field measurements did not 
see the cross over for 2/5, so it can be seen that a combination of experimental techniques is 
required for the complete study of the FQHE.

In summary we have measured the FQHE gaps at $\nu=2/3$, 2/5 and 1/3 under conditions where 
the Zeeman energy can be tuned through zero. For the ferromagnetic state at $\nu=1/3$ the 
energy gap decreased dramatically as the ZE was reduced to zero and recovered again once the 
sign of the g-factor changed. At small ZE, the excitation appears to consist of 3 reversed spins 
which we interpret as a small composite skyrmion. The behavior is very similar to that seen for 
the most easily accessible quantum Hall ferromagnet state at $\nu=1$, and is in general 
agreement with theoretical predictions. These experiments lend support to the existence of 
skyrmionic composite Fermions excitations within the two-dimensional electron gas.

{\bf Acknowledgements\\}
This work is supported by European Union TMR Programme number ERBFMGECT950077 and 
NATO Research Grant 930471.

\newpage

\begin{references}

\bibitem{fqhe} D.C.\ Tsui {\em et al.}\ Phys.\ Rev.\ Lett.\ {\bf 48}, 1559 (1982); R.L.\ Willett 
{\em et al.\ ibid.}\ {\bf 59}, 1776 (1987); T.\ Chakraborty \& P.\ Pietilainen, {\em The Quantum 
Hall Effects} (Springer-Verlag, New York, 1988); S.\ Das\ Sarma \& A.\ Pinczuk, {\em 
Perspectives in Quantum Hall Effects} (Wiley, New York, 1997) 

\bibitem{cft} J.K. Jain, Adv.\ Phys.\ {\bf 41}, 105 (1992); B.I.\ Halperin, P.A.\ Lee and N.\ 
Reed, Phys.\ Rev.\ B {\bf 47}, 7312 (1993)

\bibitem{cfe} D.R. Leadley {\em et al.}\ Phys.\ Rev.\ Lett.\ {\bf 72}, 1906 (1994); R.R.\ Du 
{\em et al.}\ Phys.\ Rev.\ Lett.\ {\bf 73}, 3274 (1994)

\bibitem{nmr} S.E.\ Barret {\em et al.}\ Phys.\ Rev.\ Lett.\ {\bf 74}, 5112 (1995); E.H.\ Aifer 
{\em et al.}\ Phys.\ Rev.\ Lett.\ {\bf 76}, 680 (1996)

\bibitem{duncan} D.K.\ Maude {\em et al.}\ Phys.\ Rev.\ Lett.\ {\bf 77}, 4604 (1996)

\bibitem{nu1} D.R.\ Leadley, to be submitted to PRL

\bibitem{eise} A.\ Schmeller {\em et al.}\ Phys.\ Rev.\ Lett.\ {\bf 75}, 4290 (1995)

\bibitem{ferr} H.A.\ Fertig {\em et al.}\ Phys.\ Rev.\ B {\bf 50}, 11018 (1994)

\bibitem{sond} S.L.\ Sondhi {\em et al.}\ Phys.\ Rev.\ B {\bf 47}, 16419 (1993)

\bibitem{dl32} R.J.Nicholas {\em et al.}\ Semicond.Sci.Technol.\ {\bf 11}, 1477 (1996)

\bibitem{du32} R.R. Du {\em et al.}\ Phys.\ Rev.\ Lett.\ {\bf 75}, 3926 (1995)

\bibitem{tilt} R.G.\ Clark {\em et al.}\ Phys.\ Rev.\ Lett.\ {\bf 62} 1536 (1989); J.P.\ Eisenstein 
{\em et al.\ ibid.}\ 1540; L.W.\ Engel {\em et al.}\ Phys.\ Rev.\ B {\bf 45}, 3418 (1992)

\bibitem{pres} N.G.\ Morawicz {\em et al.}\ Semicond.Sci.Technol.\ {\bf 8} 333 (1993); S.\ 
Holmes {\em et al.\ ibid.}\ {\bf 9}, 1549 (1994)

\bibitem{prb} D.R.\ Leadley {\em et al.}\ Phys.\ Rev.\ B, {\bf 53}, 2057 (1996)

\bibitem{cell} M.I.\ Eremets and A.N.\ Utjuzh, High-Press.\ Sci.\ Techn., {\bf 2}, 1597 (1993)

\bibitem{eta}  Some workers use the symbol $\tilde{g}$ for $\eta$. In Ref. 8 $\tilde{g}=\eta/2$.

\bibitem{das} F.C.\ Zhang \& S.\ Das\ Sarma, Phys.\ Rev.\ B {\bf 33}, 2903 (1986); T.\ 
Chakraborty {\em et al.}\ Phys.\ Rev.\ Lett.\ {\bf 57}, 130 (1986)

\bibitem{kwj} R.K.\ Kamilla, X.G.\ Wu and J.K.\ Jain, Solid\ State\ Commun.\ {\bf 99}, 289 
(1996)

\bibitem{tilt2} R.G.\ Clark {\em et al.}\ Surf.\ Sci.\ {\bf 229}, 25 (1990)

\end{references}
\end{document}